# Molecular Dynamics Simulation Study of Interaction between Model Rough Hydrophobic Surfaces.

Changsun Eun and Max L. Berkowitz*

Department of Chemistry, University of North Carolina at Chapel Hill,

Chapel Hill, NC 27599

*Abstract*

We study some aspects of hydrophobic interaction between molecular rough and flexible model surfaces. The model we use in this work is based on a model we used previously (Eun, C.; Berkowitz, M. L. *J. Phys. Chem. B* **2009**, *113*, 13222-13228), when we studied the interaction between model patches of lipid membranes. Our original model consisted of two graphene plates with attached polar headgroups; the plates were immersed in a water bath. The interaction between such plates can be considered as an example of a hydrophilic interaction. In the present work we modify our previous model by removing the charge from the zwitterionic headgroups. As a result of this procedure, the plate character changes; it becomes hydrophobic. By separating the total interaction (or potential of mean force, PMF) between plates into the direct and the water-mediated interactions we observe that the latter changes from repulsive to attractive, clearly emphasizing the important role of water as a medium. We also investigate the effect of roughness and flexibility of the headgroups on the interaction between plates and observe that roughness enhances the character of the hydrophobic interaction. The presence of a dewetting transition in a confined space between charge-removed plates confirms that the interaction between plates is strongly hydrophobic. In addition, we notice that there is a shallow local minimum in the PMF in case of charge-removed plates. We find that this minimum is associated with the configurational changes that flexible headgroups undergo, as the two plates are brought together.

## I. Introduction

A large amount of effort was devoted to understanding the effect of water on the behavior of solutes. As a result, many useful ideas and concepts emerged, including such concepts as hydrophobic or hydrophilic interactions.[1] The hydrophobic interaction is attractive and the origin of this interaction is due to water, and not to molecules themselves. This implies that to understand the nature of hydrophobic interactions we need to separate the total interaction into contributions due to water and direct interaction; the latter may induce aggregation of molecules even without the presence of water, e.g. in vacuum.[2] This strategy should be also used when one wants to understand hydrophilic interactions. Since the interaction between two particles is usually described by the potential of mean force (PMF), which is the free energy change along a certain coordinate, this PMF can be decomposed into the aforementioned two contributions, whether the particles are rigid[2] or flexible.[3]  Recently, we used this separation method of the PMF to study the hydration force, a monotonic repulsive interaction acting between lipid bilayers[4-6]. For this purpose we performed simulations on a simplified model of lipid bilayers. To imitate the bilayer leaflets in our model, we attached the hydrophilic phospholipid headgroups to a graphene plate.[3] Our simulations demonstrated the existence of water mediated repulsive contribution into hydrophilic interactions.  Other studies also illustrated that water-mediated contribution to the typical hydrophilic interaction is repulsive[7,8].

   In our previous study we found that the origin of the hydration force is due to a strong attraction between water and charges on the bilayer headgroups[9].  Thus, it is reasonable to consider that the polarity of the headgroups is a key element in determining the character of water-mediated interaction. In this paper, we explicitly test this idea by removing charges from the original hydrophilic headgroups, and by comparing the original water-mediated interaction with the one obtained in the charge-removed case. In fact, a similar idea of changing polarity by adjusting the magnitude of charges has been used to study the effect of surface polarity on water contact angle and interfacial water structure[10]; however, not on the inter-surface interaction. We observe that the interaction between plates with removed charges on the attached headgroups is hydrophobic.

   Understanding the nature of hydrophobic interactions is now a subject of an intense activity[11-15]. In some studies the hydrophobic interaction between nanoparticles was investigated by considering the interaction between two plates containing "carbon" atoms in graphene geometry that interact across water.[16,17] It was observed that water-mediated interaction is attractive when "carbon"-water interaction is purely repulsive and that it does not change with some addition of a weak attraction to the "carbon"-water interaction. Moderate strength attraction between "carbon" and water can change the character of the water mediated interaction substantially: the PMF may start displaying spatial oscillations and the interaction between plates may display characteristics of hydrophilic interaction.[17]

Besides polarity, other factors could be important in characterizing hydrophobic/hydrophilic properties; such as density and arrangement of polar/non-polar particles and surface morphology and/or surface roughness. It is now well-established that roughness enhances the hydrophobicity of surfaces.[18-24] To understand molecular details related to the description of rough hydrophobic surface wetting, some molecular dynamics (MD) simulations were performed.[20,25-29] Most of these studies have been focused on dewetting at a single surface, in particular, calculating the contact angle of a water nanodroplet on the surface.[20,25-29]. Moreover, the model surfaces used in those studies were rigid and displayed no flexibility. To the best of our knowledge, the effect of roughness on the hydrophobic interaction has rarely been discussed in a context of water properties in the confined space between two rough surfaces, i.e. in terms of inter-surface interaction and dewetting transition. In many situations, especially in a biological environment, roughness of hydrophobic surfaces, such as protein surfaces, may influence the hydrophobic interaction between surfaces. In addition, many materials in nature are soft, so their surfaces are flexible. Thus, by investigating hydrophobic interactions between our charge-removed PC-headgroup plates we study the interactions between model rough and flexible surfaces in water.

Our paper is organized in the following way: In the Methodology section we describe the details of our systems and of our MD simulations. In the section after that we discuss the effect of the headgroup polarity on the water-mediated interaction between two model lipid bilayers and directly compare hydrophilic and hydrophobic interactions observed to act between plates with charged and uncharged headgroups, correspondingly. After that we study the hydrophobic behavior of our charge-removed model and compare it to the behavior obtained for interactions between smooth hydrophobic plates. In the final section we summarize our findings with the conclusions.

## II. Methodology

To discuss the water-mediated interaction acting between two hydrophilic surfaces, we have used the results we obtained from our previous study of the hydration force.[3] In that study, a model hydrophilic surface contained nine polar phosphatidylcholine (PC) lipid headgroups that were attached to a nanoscale graphene plate (252 carbon atoms; 2.425 nm by 2.380 nm). We named the plate the PC-headgroup plate and the details about this PC-headgroup plate are given in our previous paper.[3] In Figure 1(a) we show the PC-headgroup plates and in Figure 1(b) the charge distribution on the PC headgroup. To study water-mediated hydrophobic interaction between rough and flexible surfaces, we constructed a hydrophobic plate closely related to the original PC-headgroup plate. This hydrophobic plate, named the charge-removed PC-headgroup plate (CRPC plate), was prepared by removing all the electric charges from the lipid headgroups in the original PC-headgroup plate. Therefore, the interaction between the plate and water in this case was just the Lennard-Jones interaction.

To calculate the water-mediated interaction for the CRPC plates, we performed MD simulations and employed the same methodology we used in our previous work.[3] Thus we calculated the potential of mean force (PMF) as a function of the interplate distance by using thermodynamic perturbation method.[16,30] Again, as in the previous work, the interplate distance is defined as the distance from one graphene plate to the other graphene plate. As previously, we decomposed the PMF into direct and water-mediated contributions using the procedure described in the Appendix in reference 3. The errors in the calculations were estimated using the block averaging method[31] and the standard error propagation method [32,33].

In our MD simulations of the system with the CRPC plates, we placed the plates into a simulation box separated by a certain interplate distance and solvated them with 8800 water molecules. We used the NPT ensemble; the Nose-Hoover temperature[34,35] and the Parrinello-Rahman pressure[36] coupling algorithms (both with a coupling constant of 0.5 ps) were utilized for maintaining temperature at 298 K and pressure at 1 bar, respectively. Electrostatic interaction was calculated through the particle mesh Ewald method.[37] The SPC/E model[38] was used for water. Periodic boundary conditions were employed. The same interaction and the same simulation parameters as we used in our previous study[3] were also used, except for the length of the simulation time. That is, when the interplate distances were in a region between 1.68 nm and 1.82 nm the total simulation time for each distance was 10 ns, to permit enough sampling, because we observed that a dewetting transition is happening inside that region; otherwise the time of simulation runs was 1 ns. In our calculations of the PMF and other physical quantities the data from the first 500 ps were discarded.

To characterize the hydrophobic interaction between CRPC plates, we compared the calculated PMF and its components with the calculations we performed previously on a series of cases with simple graphene-like "carbon" plates that had no flexible headgroups attached.[17] In reference 17 we considered a series of simulations with "carbon" plates in graphene geometry, where every simulation differed from the other by the strength of the Lennard-Jones (LJ) potential acting between the "carbon" of the plate and water.[17,39] Seven systems where the LJ interaction was moderately strong or weak were investigated. For these systems the water- "carbon" interaction strength was determined through the use of the Lorentz-Berthelot combination rules for the LJ parameters of "carbon"-"carbon" and oxygen-oxygen interaction, i.e. $\varepsilon_{CO} = \sqrt{\varepsilon_{CC} \cdot \varepsilon_{OO}}$ and $\sigma_{CO} = (\sigma_{CC} + \sigma_{OO})/2$. As was mentioned previously[17], we fixed the values of $\varepsilon_{OO}$ and $\sigma_{OO}$, taken from the SPC/E water model, and $\sigma_{CC} = 3.4\text{Å}$. Note that for a realistic graphene plate, the LJ parameter $\varepsilon_{CC}$ is 0.3598 kJ/mol[16], for the other cases we chose $\varepsilon_{CC}$ values to be 0.05, 0.10, 0.15, 0.20 and 1.00 kJ/mol. In addition, we considered the case in which the LJ potential was purely repulsive as a limiting hydrophobic case. For this, we used Weeks-Chandler-Andersen (WCA) separation scheme[40] based on $\varepsilon_{CC}$ =0.3598 kJ/mol.

For all types of "carbon" plates the reference state for the PMF (the state where the value of the PMF is assumed to be zero) was at an interplate distance of 1.40 nm, whereas the reference state for the PMF of PC-headgroup and CRPC plates was at 2.99 nm. The simulation time for systems with "carbon" plates for each interplate distance was 2 ns. In these cases we also disregarded the first 500 ps of trajectories for the data analysis. For the PMF calculations in "carbon" plate systems we used the 60-atom graphene-like plate as Choudhury and Pettitt did.[16]

All MD simulations were carried out using the GROMACS 3.3.1 and 3.3.3 packages.[41]

### III. Discussion

**Role of the electrostatic interaction in a hydrophilic interaction**

From our previous study of the interaction between model lipid bilayer plates we concluded that the repulsion between neutral lipid bilayers originates from the water-mediated interaction[3] (see Figure 2(a)), and that this interaction is mainly due to the increase in the potential energy of interaction between the model lipid plates and water molecules, as the two plates are brought together.[9]    In addition, the hydrogen bonding analysis showed that this potential energy change is inversely correlated with the change in the number of hydrogen bonds between the plates and water.[9] All this implies that electrostatic interaction between the polar headgroups and water plays an important role in the origin of the repulsive hydration force.

   To find out how much the electrostatic interaction contributes to the total potential energy of interaction between the plates and water, we calculated the electrostatic and the LJ interaction contributions to the water-plates interaction energy. Since we are interested in the change of energy as the plates are brought together from large separations, we focused on the energy difference $\Delta U(r) = U(r) - U(r = 2.99nm)$, where r is the distance between the plates and U is the water-plate interaction energy. We display the electrostatic and LJ components of this energy difference in Figure 3. As we can see from this Figure, the energy change due to electrostatic interaction is dominant. This confirms our suggestion that the electrostatic interaction between polar headgroups and water molecules is so strong, that it requires work to remove water molecules when the two model plates are brought together.

**Hydrophilic interaction vs. hydrophobic interaction**

Since the electrostatic interaction is crucial for generating repulsive interactions between PC-headgroup plates, we expect that upon removal of the water-headgroup electrostatic interaction, the repulsive character of interbilayer interaction would disappear. To test this idea, we use our PC-headgroup plate again, but this time we remove all electric charges from the headgroups attached to plates and then calculate the interplate interaction. Obviously, since the electrostatic interaction between the plate and water is absent now, this weakens the water-plate interaction compared to the original hydrophilic plate. As a result, the PMF for the interplate interaction between the CRPC plates is not repulsive, but attractive (see Figure 2(b)). From the decomposition of the PMF we observe that, while in case of plates with charged headgroups the repulsive contribution due to water into the PMF was determining the total repulsive character of the PMF, the attractive water-mediated interaction in case of CRPC plates is contributing to the attractive total interaction between plates. Since the water-mediated interaction is attractive for hydrophobic interactions, we can consider the interaction between our CRPC plates to be hydrophobic. Indeed, the overall shape of the PMF in Figure 2(b) looks like the shape observed in other cases with hydrophobic interaction. [16,42,43]

Besides performing the energetic analysis, we also examined the character of the dewetting transition (when the number of water molecules in the confined space between the plates is reduced abruptly over a certain distance interval) in the charge-removed case. Such dewetting transition is considered to be another signature of a hydrophobic interaction.[13] In order to do that, we calculated the average number of water molecules confined between the two CRPC plates. Calculations of water density and also direct observation of location of water molecules in a region between the plates show that some of the water molecules are always present at the edges of the plates. To avoid the contribution of edge effects, we calculated the number of water molecules in a rectangular prism located in the confined space between the CRPC plates. The base of the prism was taken to be a 0.8x0.8 nm$^2$ square centered at the center of our plate.. The result for the number of water molecules is shown in Figure 4(a). It suggests that the dewetting transition actually occurs in the shaded narrow region (d=1.74~1.8 nm). In addition, the changes in an instantaneous number of water molecules in the confined space and the snapshots from the simulations provide us with more clear evidence of the transition existence, as Figures 4(b) and 4(c) show. We observe from Figure 4(c) that a large cavity forms inside the interplate space, and the thickness of the cavity along the axis perpendicular to the plates is large enough to accommodate approximately two layers of water molecules.

***"Carbon" plates as reference hydrophobic plates.***

Our CRPC plates containing flexible non-polar headgroups can be considered as rough and flexible hydrophobic surfaces. To understand better the degree of hydrophobicity of the CRPC plates, we compare the system containing two CRPC plates with the system containing two

"carbon" plates that represent smooth and rigid surfaces. The reason why we choose systems with "carbon" plates, including graphene plates, as our reference systems is that they have been relatively well-studied.[16,39,44] Thus, a comparison study can help us to understand the common features characteristic for plates that interact through hydrophobic interactions and distinct features due to roughness and flexibility of surfaces.

Besides a graphene plate, for a systematic comparison, we considered six other "carbon" plates, which have the same geometry as a graphene plate, but different interaction strength for the water-plate interaction. The graphene plate we considered is not a typical hydrophobic plate since the contact angle for water with such a plate is around 90° [45] and the water-mediated interaction between such plates is not purely attractive[16,17]; moreover, the dewetting transition for water between graphene plates is hardly observable.[17] The way we adjust the interaction strength for water-plate interaction, or the plate hydrophobicity, is through variation in the value of $\varepsilon_{CC}$, as described in the Methodology section. It was shown by Choudhury and Pettitt[39] that it is possible to induce a pronounced dewetting transition for water between two graphene-like plates by reducing the LJ parameter of "carbon"-water interaction. They performed their study only at a particular interplate distance of $6.8\,\text{Å}$, but it showed that by controlling water-plate interaction we may also control the strength of hydrophobic interaction.

To understand the dependence of the hydrophobic interaction on water-"carbon" interaction strength, as well as the character of the dewetting transition, we previously performed MD simulations for a certain range of water-plate interactions and for a range of interplate distances.[17] Figure 5(a) shows the PMFs and the average numbers of water molecules in the confined space for different strengths of "carbon"-"carbon" interaction. From the figure we see that as the value of $\varepsilon_{CC}$ decreases from the value of 1.00 kJ/mol to 0.05 kJ/mol, the water-mediated interaction becomes more attractive, indicating that a stronger hydrophobic interaction is acting between plates. Also, the reduction of water-plate interaction strength affects the behavior of the average number of water molecules. That is, the dewetting transition regions, characterized by the large changes in the number of water molecules (left shaded region in Figure 5(a)) are getting broader and they shift towards larger distances. When the "carbon"-water interaction is reduced (this corresponds to $\varepsilon_{cc}$ values of 0.10 and 0.05 kJ/mol) the dewetting region and the transition region between one and two layers of water (right shaded region in Figure 5a) are merged into one region. For pure repulsive potential (WCA) case, the transition occurs at much larger interplate distances. Overall, the analysis of the average number of water molecules shows that as "carbon"-water interaction decreases, the character of hydrophobic interaction between "carbon" plates increases and the dewetting transition region gets shifted towards larger interplate distances.

***Hydrophobicity of the charge-removed PC-headgroup plate***

For an appropriate comparison between the CRPC plates and reference "carbon" plates we recalculated the distance between the CRPC plates as the distance between opposing choline groups. This distance represents a better choice, since, as we observed, water does not penetrate the uncharged headgroups. Therefore, most of the headgroup, except cholines, can be considered to be inside of the CRPC plate. Thus, the length of the fluid space between the CRPC plates is the intercholine distance, rather than the graphene-graphene or interplate distance. Since the headgroups extend from the plate by a distance around 0.5 nm, roughly the distance between headgroup planes (the size of the fluid space between plates) should be shorter by 1 nm compared to the interplate distance. More careful measurement shows that the relationship between the interplate distance (x, in nm) and the distance between opposing planes determined by the centers of mass of choline groups (y, in nm) is $y = 1.046x - 1.034$ (the correlation coefficient of 0.99). Using this relationship we present in Figure 5(b) the PMF and the number of confined water molecules as a function of the intercholine distance.

From the comparison of Figures 5(a) and 5(b) we notice that the interaction between CRPC plates is similar to the interaction between "carbon" plates with a weak "carbon"-water interaction ($\varepsilon_{CC}$ =0.05 kJ/mol). To investigate the similarity furthermore, we estimated the value for the average Lennard-Jones potential energy of interaction between a plate and a single water molecule (Figure 6). For this we performed a series of NVT simulations on a system containing 252-atom "carbon" plate, so that its size is exactly the same, as the one used in the simulations of the CRPC plates. In the case of the CRPC plate we took an average of the LJ interaction over the 200 ps trajectory keeping water at a given distance, because the headgroups are fluctuating and so is the LJ interaction. In our calculations the plate-water distance along the z-axis (perpendicular to the plate) in case of a "carbon" plate was defined as the distance between the center of mass of the plate and the water molecule, while for the distance between the CRPC plate and water it was the average distance between the choline group and the water along the z-axis (see Figure 6(a)). Since the surfaces of plates are inhomogeneous on the atomic scale, we considered five different x-y positions of the water molecule (see Figure 6(b)) and calculated the average of the LJ interactions over these configurations (Figures 6(c) and 6(d)). The results show that when a water molecule approaches the CRPC plate, it is weakly interacting with the plate. As a matter of fact, up to the comparable distance (~ 0.34 nm), the behavior of the energy curve for the CRPC-water interaction is very similar to the behavior of the curve for the LJ interaction between "carbon" plates and water, when "carbon" plates interact weakly with water ($\varepsilon_{CC}$ =0.05 and 0.10 kJ/mol), as the Figure 6(d) shows. This result is consistent with the results obtained from consideration of PMFs and dewetting transitions.

Further confirmation that our CRPC plates are similar to plates where "carbon"-water interaction is weak is obtained from considering the snapshots of water and calculating the water density profiles, especially in a region that is close to the dewetting transition. The density profiles are shown in Figure 7(a). Note that the change of density profiles in the dewetting transition regime (d=1.74~1.8 nm) is correlated with the large change in the average number of water molecules depicted in Figure 4(a). From Figure 7(b) we observe that water between the graphene plates (black curve, $\varepsilon_{CC}$ =0.3598 kJ/mol) shows a strong layering structure, as the distance between plates decreases. However, water between the CRPC plates does not show layering. Water between "carbon" plates with $\varepsilon_{CC}$ =0.05 kJ/mol and between the CRPC plates show a similar behavior in the dewetting transition regime; compare the changes depicted in Figure 7a and the changes of profiles in the confined space from Figure 7(b) when d ≤ 0.9 nm.

### *Role of roughness*

Let us now discuss the effects of roughness on the hydrophobic attraction. Firstly, let us discuss how representative our CRPC plate is, since the headgroups of this plate are artificial and are composed of non-polar atoms derived from lipid atoms. Although these atoms are not realistic, their LJ values are in the range of values for the elementary atoms defined in GROMACS G43a1 force field. Thus, our CRPC plate is representative of a non-polar moiety. Interestingly, if we replace every P, O and N atom in artificial headgroup by a proper hydrocarbon group, we observe that in our CRPC plate the atoms underestimate the hard-core radius and overestimate dispersive attraction compared to a corresponding plate with hydrocarbon groups. This implies that the CRPC place is probably more hydrophilic than the corresponding hydrocarbon plate.

To understand why the plate with attached hydrophobic headgroups that create a rough surface is more hydrophobic compared to the smooth graphene plate, let us consider Figure 8, where we compare water density between graphene plates and water density between our model CRPC plates. We chose 1.86 nm for the separation distance between the CRPC plates because at this distance the system is in a wetting state. For an appropriate comparison we again used the relationship between distances x and y ($y = 1.046x - 1.034$), as explained above, so that the average positions of cholines of the CPRC plates are corresponding to the positions of the graphene plates. For this particular distance, we calculated the density profiles of water, cholines and phosphates. This result is shown in Figure 8(a) and the result for the corresponding graphene case is shown in Figure 8(b).

In the case of the CRPC plate the confined water molecules interact with headgroups only. As one can see in Figure 8(a), the water molecules are too far away to interact with the graphene part of the CRPC plates. However, in the absence of hydrophobic headgroups, the confined water molecules can interact with the graphene plate, (interaction indicated by a white arrow in the Figure 8(b). As a result, in case of the CPRC plate, water molecules interact with the low-density protruded hydrophobic parts of the headgroups, while water interacts with the high-carbon density smooth surface in the case of the graphene plate. This, consequently, reduces the water-CRPC plate interaction, compared to the graphene plate and, therefore, the interaction between the CRPC plates has a strongly hydrophobic character.

***Role of flexibility due to non-polar headgroups***

Since the length of a headgroup (composed of 11 atoms and united atoms) is relatively small compared to the graphene plate size and the one end atom of the headgroup is fixed because it is attached to the plate, the effect of the flexibility is relatively small. However, we are able to capture this effect in the PMF curve. Specifically, it is clearly seen in the shape of the curve for the direct interaction contribution to the PMF. As we can see from Figure 4a, the PMF for the direct interaction between the CRPC plates has, besides the global minimum, a local minimum around d=1.5 nm, while the PMF for the direct interaction between graphene plates does not have such a minimum (see open circles in Figure 5a). To understand the existence of this minimum, we looked at some snapshots from the simulations. These snapshots, shown in Figure 9a, indicate that the minimum in the direct interaction part of the PMF curve is associated with the change of relative orientations of the headgroups protruding from one plate with respect to the headgroups from the other plate.

For a quantitative analysis of this reorientational motion we calculated a xy-dimensional pair correlation function between the two groups of headgroups (specifically, methyl groups of choline moieties; blue and red spheres in Figure 9(a)). To calculate this correlation function we used the GROMACS program utilities and the result of the calculations is shown in Figure 9(b). It is clear that there is a conformational change when the distance between the CRPC plates is in the interval between 1.3 nm and 1.5 nm. Note that at these distances the partial dewetting already took place and water does not play a significant role. At larger separations (d > ~ 1.5 nm), the preferable configuration is such that blue and red headgroups overlap in the xy plane, to maximize the van der Waals interaction between them. Also, for the same reason, the headgroups in a dewetted state are a little bit stretched along the z axis (not shown here). However, when the headgroups are close enough (d < ~ 1.3 nm), such configuration is not preferable any more, due to steric repulsion. Thus, in the regime of the intermediate separations, the headgroups change their relative orientations, so they can avoid repulsion acting between them. As a result, the overlaps in the xy plane disappear. This is explained using cartoons in Figure 9(c). Thus, due to flexibility of the headgroups, the system can reach a deeper minimum in the direct part of the PMF.

### IV. Summary and Conclusions

We observed that electrostatic interactions play an important role in the nature of the repulsive hydration force acting between model lipid bilayers we called PC-headgroup plates and that it is crucial for understanding the thermodynamic origin of the hydration force.[9] To present further evidence justifying this conclusion, we removed the electric charges from headgroup atoms and calculated the free energy as a function of distance between the two, now non-polar, model bilayers (or CRPC plates). Contrary to the case of the original model bilayers, the interlayer interaction became attractive. This again supports the notion that the repulsive interaction between the PC-headgroup plates originated from the electrostatic interaction between polar headgroups and water molecules.

To understand how hydrophobic the interaction between our CRPC plates is, we compared the results from the simulations with these plates to the results obtained from simulations where interaction between smooth plates containing "carbon" atoms with a variable degree of "carbon"-water interaction. The comparison showed that our CRPC plates are strongly hydrophobic. Attachment of hydrophobic groups to smooth plates increased the plate hydrophobicity, because the groups created voids between water and graphene surface and water molecules were not able to fill up these voids. Therefore the state of water in our simulations with the CRPC plates was similar to the Cassie-Baxter state,[19] and this contributed to the increase of the hydrophobicity of the interplate interaction. The flexibility of the plate headgroups also influenced the interaction, mostly the direct interaction between the plates.

In this study and our previous studies on this subject,[3,9,17] we did not calculate the contact angle between the water droplet and our plates, to determine the hydrophobic (hydrophilic) character of the plate. We did not do this, because our plates are small, and it is very hard to calculate accurately the microscopic contact angle of a water nanodroplet. However, when we looked at the snapshots from the simulations of systems containing drops of 300 water molecules placed on different plates (not shown here), we observed that water nanodroplet on the "carbon" plate beads up or reduces its spread over the surface, as the "carbon"-water interaction decreases. The shape of the water droplet on the CRPC plate is similar to the shape appearing in the cases of strong hydrophobic "carbon" plates.

Finally, we would like to mention that the systems we studied here and previously,[3,9,17] especially the system containing the CRPC plates, may seem to be somewhat artificial. Nevertheless, by performing a systematic study on different plate systems and their hierarchy (graphene plates, "carbon" plates with different water-"carbon" interaction strength, dressing up the plates with zwitterionic headgroups, removing charges on these headgroups) allowed us to understand the role of important factors in the phenomena of hydrophobic and hydrophilic interactions.

**Acknowledgement:** This work was supported by a grant N000141010096 from the Office of Naval Research.

**Figure Captions:**

**Figure 1.** (a) A snapshot of the PC-headgroup plates in water at an interplate distance of 2.4 nm. (b) Detailed structure of the PC-headgroup. The numbers in parentheses represent the magnitudes of partial charges (in units of the elementary charge, $e$).

**Figure 2.** Decomposition of the PMF (black curve) into contributions from the direct interaction (red) and water-mediated interaction (green) between two PC-headgroup plates (a) and between two charge-removed PC-headgroup plate (b). Insets are for large interplate distances. Errors are represented by bars.

**Figure 3.** Contributions of electrostatic (red) and Lennard-Jones (green) interactions to the potential energy interaction between the PC-headgroup plates and water molecules (black). The energies were directly calculated from MD simulations. The error bars represent the standard deviations.

**Figure 4.** (a) Average number of water molecules in the confined space (see text) between two CRPC plates, the total PMF, the contribution of direct interaction into the PMF, and the contribution of water-mediated interaction. (b) Changes in the number of water molecules in the confined space as functions of time at the distances of 1.82 nm (blue), 1.79 nm (red), 1.74 nm (black) and 1.7 nm (green). (c) For the case of an interplate distance of 1.79 nm, two snapshots taken at t = 3500ps (left) and t = 7800ps (right). Carbon and united carbon with hydrogen, nitrogen, phosphorus, and oxygen are colored in cyan, blue, tan, and red, respectively. For clarity water molecules are represented by yellow. To eliminate edge effects in the calculation of water molecules in (a) and (b), we take into account only those water molecules located in the rectangular prism between plates. The volume of the prism is a product of the square (white lines in (c)) area times interplate distance (see text).

**Figure 5.** (a) PMF (black filled square), direct interaction (black open square), and water-mediated interaction (red triangle) of "carbon" (graphene-like) plates with different water-plate interaction strength. The number in each panel represents $\varepsilon_{CC}$. The blue squares curves are for the average number of water molecules in the confined space. (b) The PMF, its components and the number of water molecules as a function of distance between CRPC plates. The interplate distance is replaced by the intercholine distance. For proper comparison, the number of water molecules is rescaled by multiplying the number of water molecules in Figure 4 (a) by a scaling factor of 1.91. This number is obtained from the requirement that the area of the rectangle base from Figure 4 (c) is equal to the area of the "carbon" plate.

**Figure 6.** Plate-water Lennard-Jones (LJ) interaction. For a direct comparison with the CRPC plate, we increased the plate size of "carbon" plates so that their area is the same as the area of the CRPC plate, which is 4.7 times larger than the "carbon" plate used in Figure 5a. (a) An initial configuration at a water-plate distance of 1.01 nm, in which a water molecule is placed on the top of the center of the plate. (b) Five different positions for sampling in case of the CRPC plate. The same sampling was also applied to each case of "carbon" plate. (c) Plate-water LJ interaction of the CRPC plate as a function of plate-water distance. The error bars are obtained by calculating the standard deviation from five samples. (d) Plate-water LJ interactions as a function of the distance between the plate and a water molecule for the "carbon" plates, and as a function of distance between the cholines and a water molecule for the CRPC plate. Inset is for showing plate-water LJ interaction for smaller distances in case of the CRPC plate.

**Figure 7.** Water number density profiles for the systems containing the CRPC plates (a) and graphene plates (b). Only water molecules from the central rectangular region of plates (see Figure 4c) are considered for (a), while the whole region between plates is considered for (b). The density profiles are normalized so that the values in bulk water region are 1 and the z coordinate of the middle point in the confined space is set to zero.

**Figure 8.** Profiles for the number of water molecules represented by oxygen atoms (black), the center of mass (COM) of three end carbons of cholines (red), and the COM of phosphates (green) along the z axis. (a) The system with CRPC plates at an interplate distance of 1.86 nm. (b) The system with graphene plates at an interplate distance of 0.91 nm. The profiles associated with cholines and phosphates are normalized, so that the maximum values are 1, whereas for water the curves are normalized by the number density of bulk water. Note that in (a) the plates are located at both ends of the z axis in the plot, and in (b) they are located where the dashed blue lines are. The white arrows represent the interaction between water and headgroups in (a) and the interaction between water and graphene plates in (b). The length scale of the white arrow is ~0.35 nm, which is approximately the value of $\sigma_{co}$

**Figure 9.** (a) Snapshots taken at 1 ns for some selected interplate distances. The perspective is perpendicular to the graphene plates. United carbon atoms (methyl groups of choline moieties) are represented by van der Waals spheres. To distinguish two types of the united atoms, depending to which plate they belong, we use red and blue colors. The graphene plates are parallel to the paper. (b) xy-dimensional radial distribution functions of the red united carbon atoms of the bottom plate, with respect to the blue united carbon atom of the top plate. (c) Schematic diagrams for explaining why the direct interaction has a small barrier between1.3 nm and 1.5 nm

**(a)**

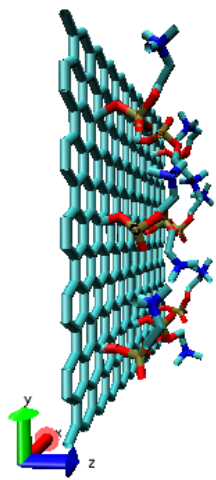

**(b)**

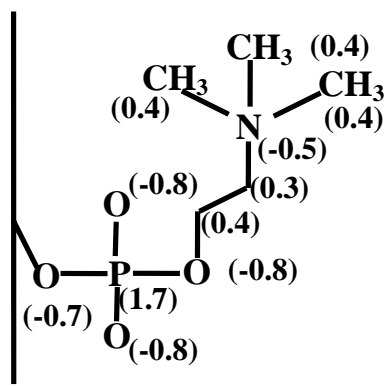

**Figure 1.**

**(a)**

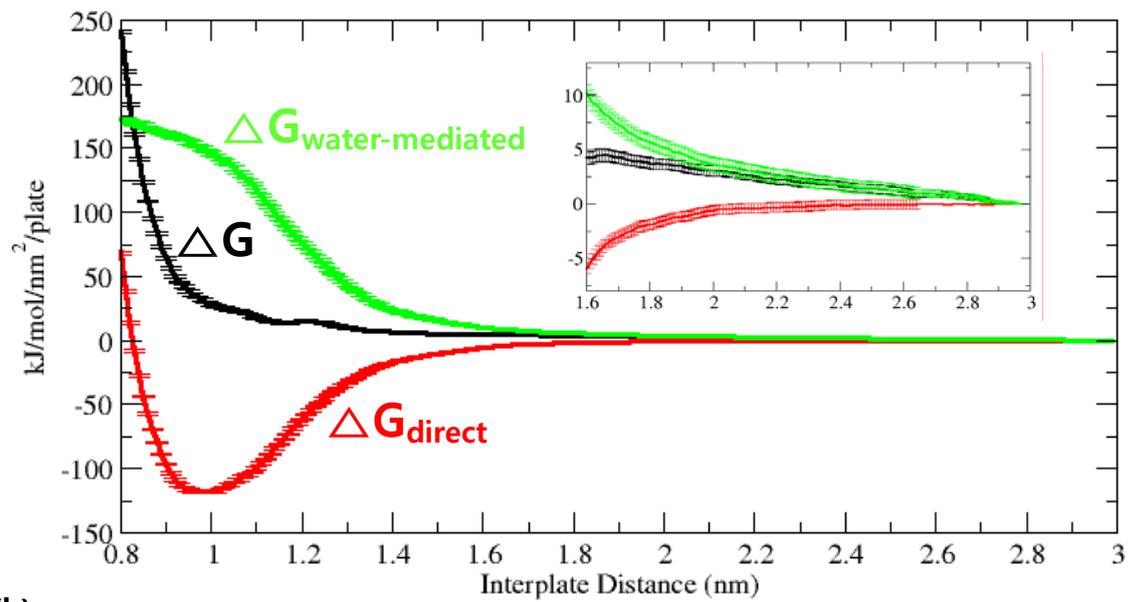

**(b)**

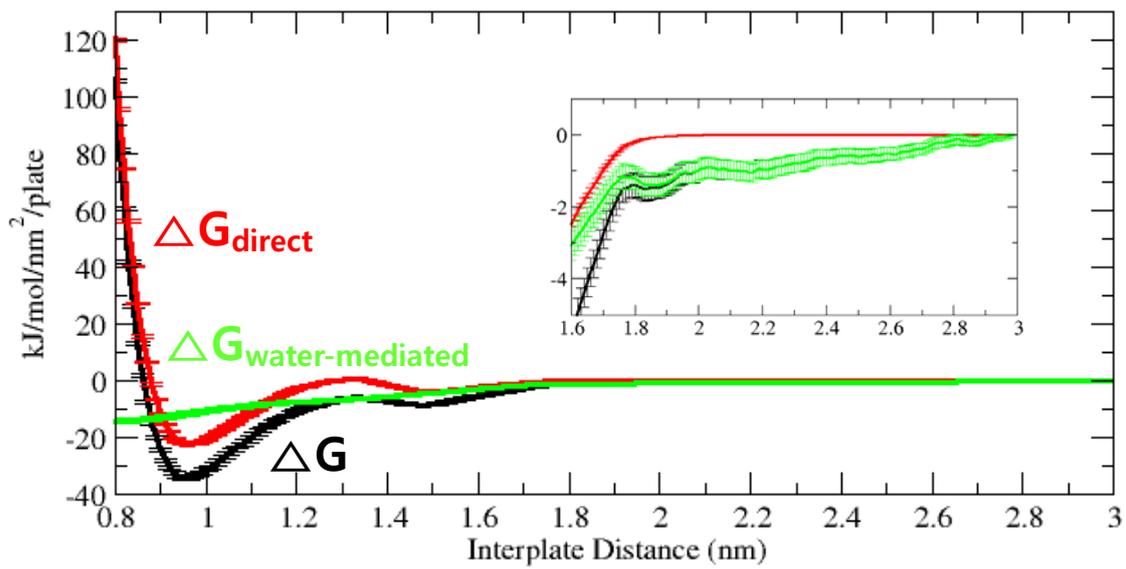

**Figure 2.**

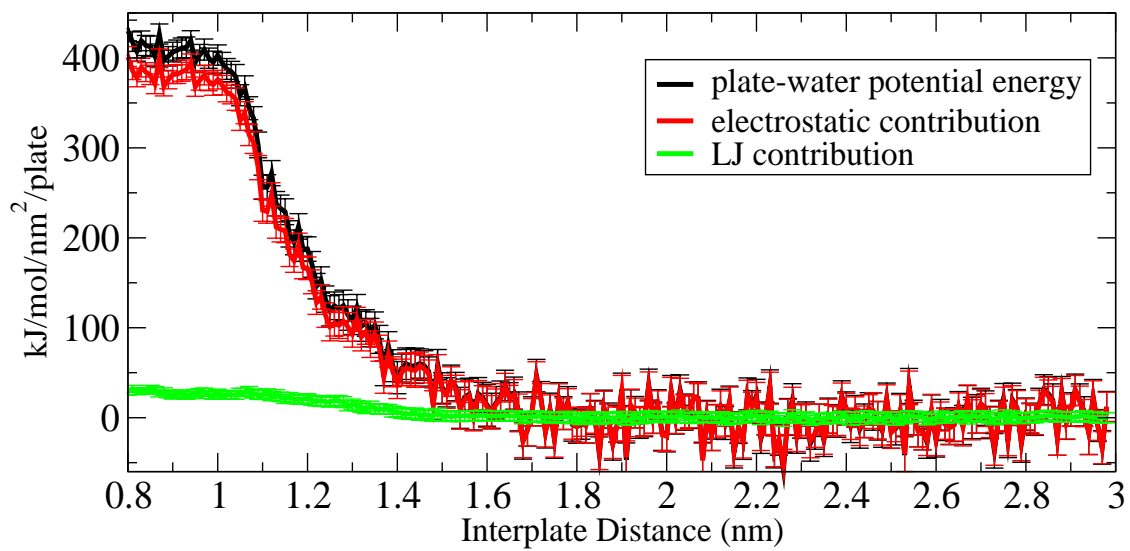

**Figure 3.**

**(a)**

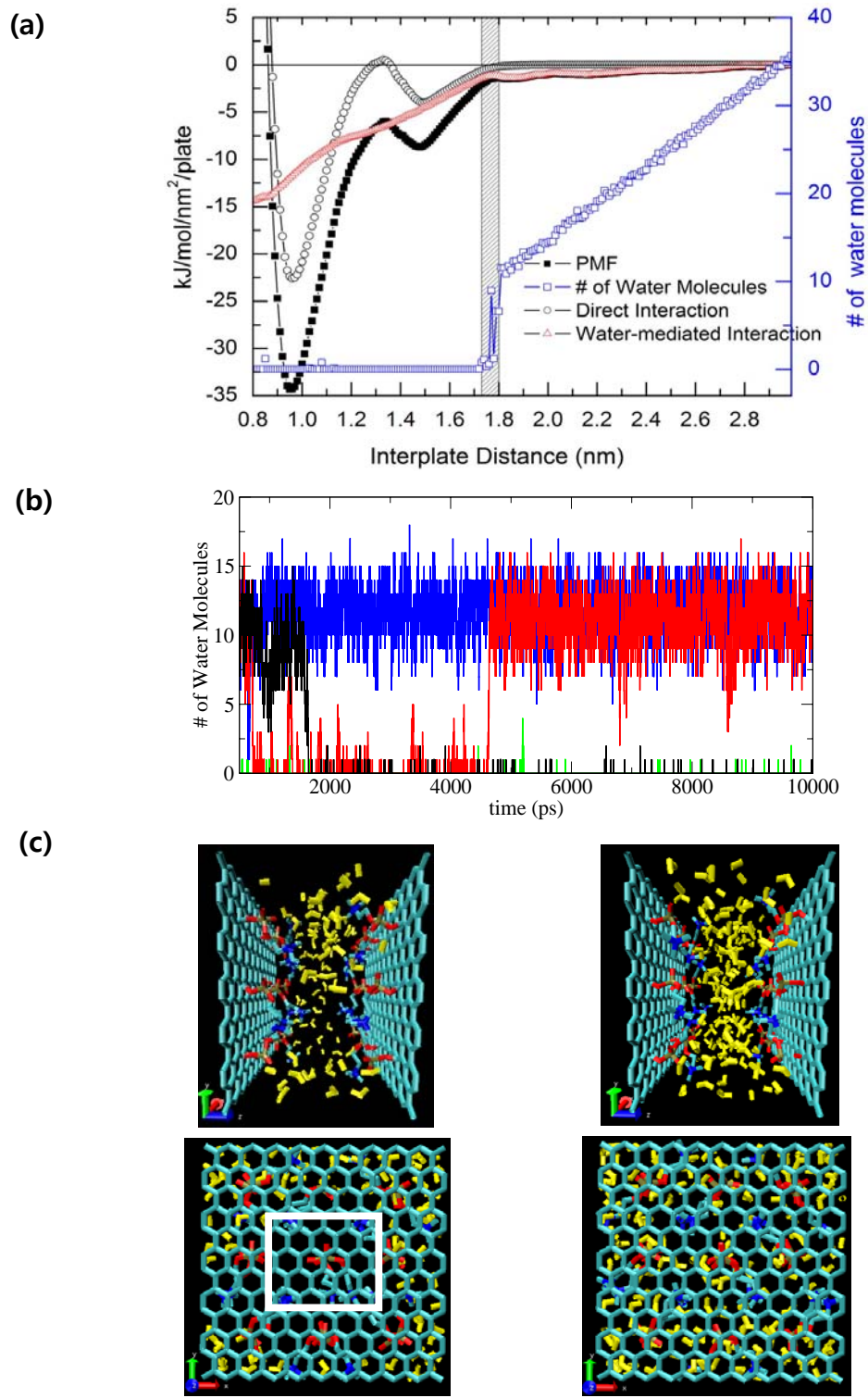

**(b)**

**(c)**

**Figure 4.**

**(a)**

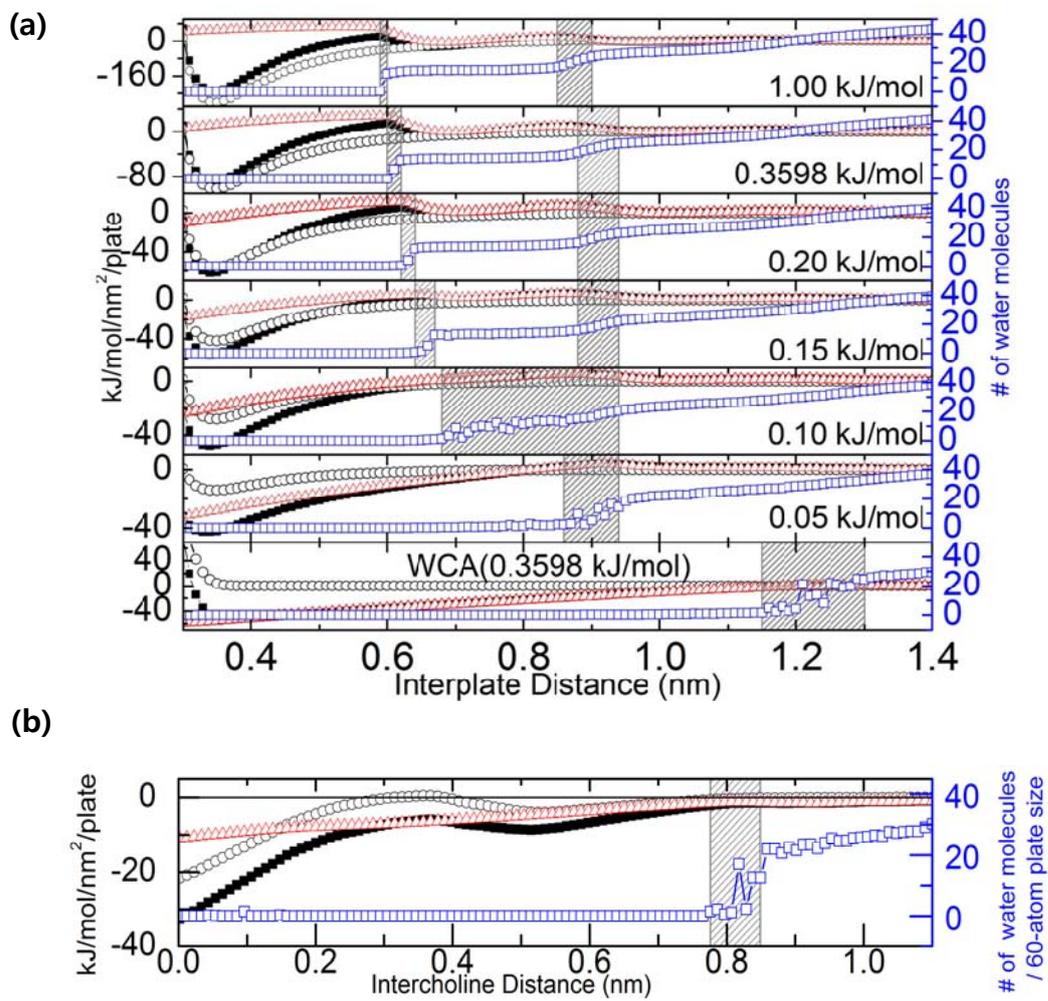

**(b)**

**Figure 5.**

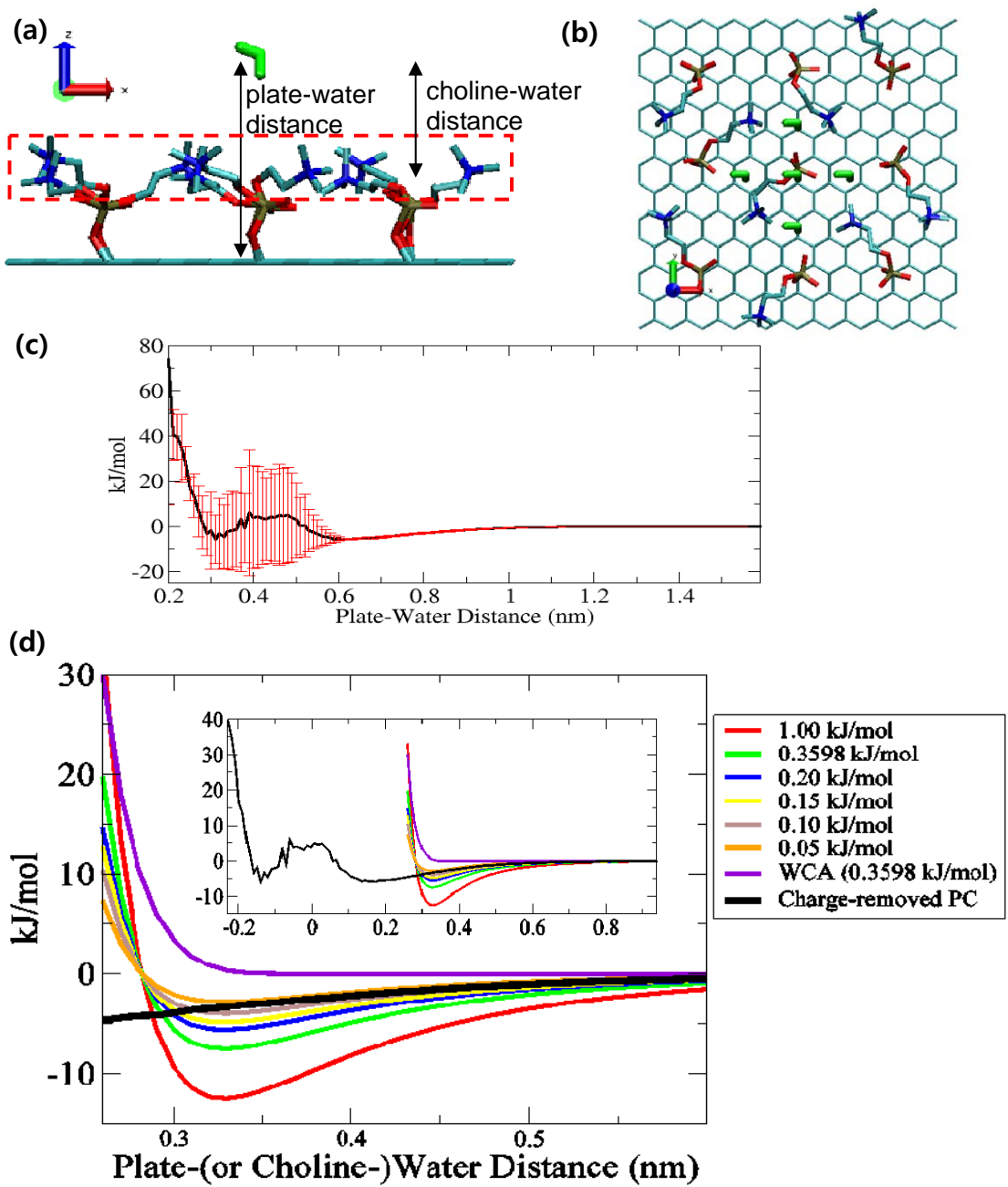

**Figure 6.**

**(a)**

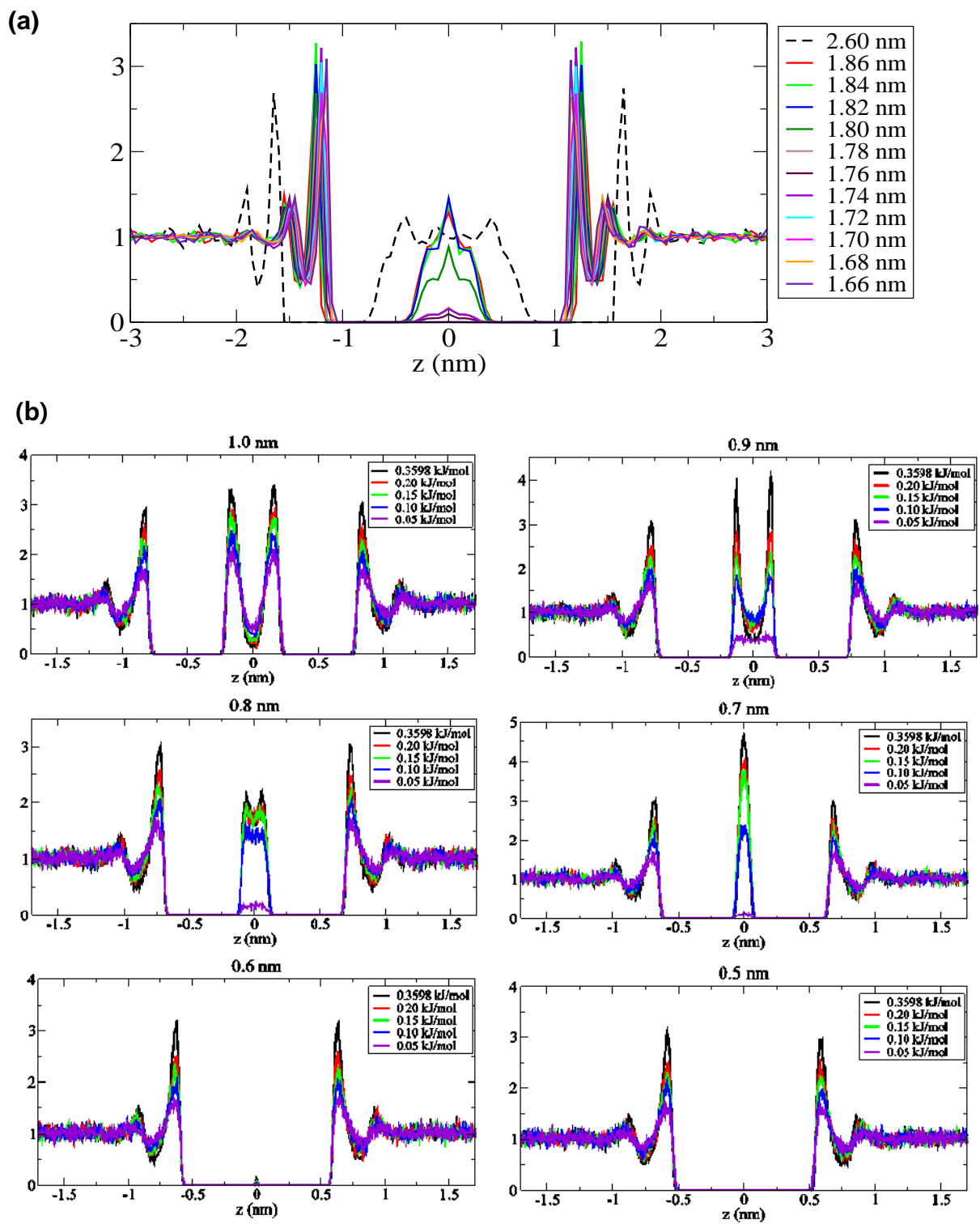

**(b)**

**Figure 7.**

**(a) CRPC plate, d=1.86 nm**

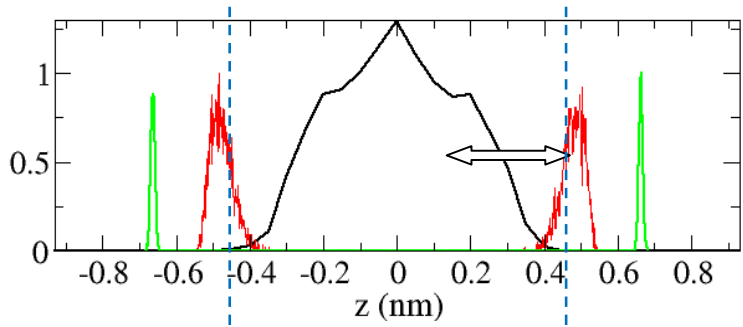

**(b) graphene, d=0.91 nm**

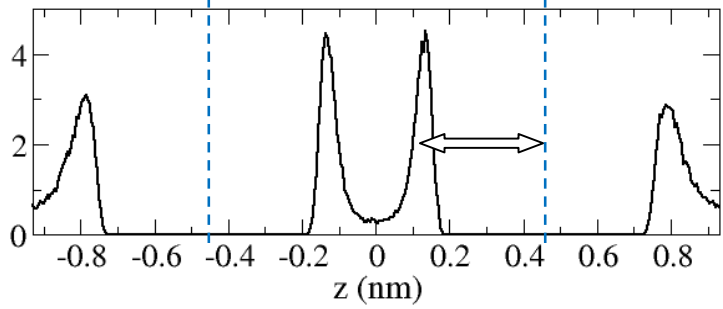

**Figure 8.**

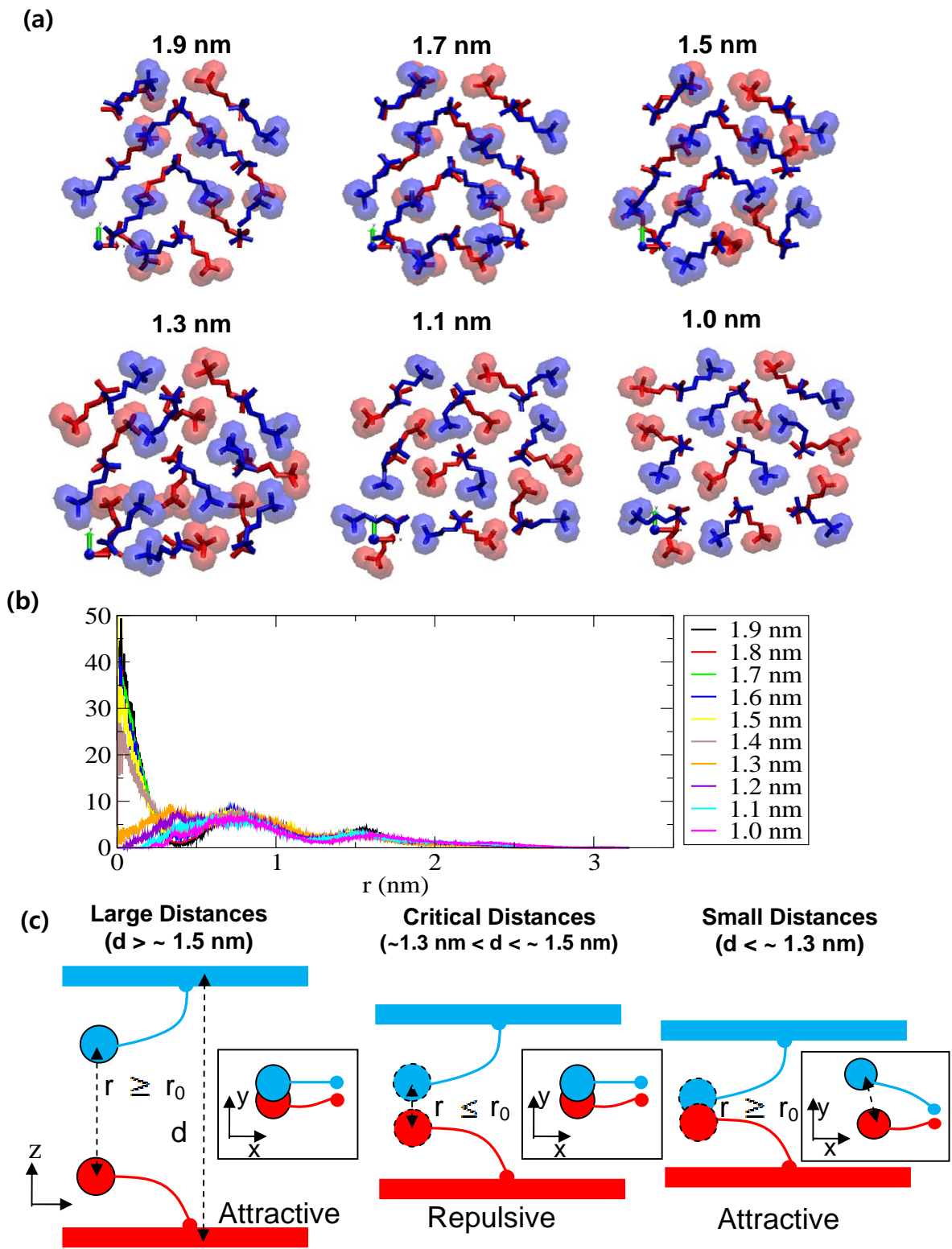

**(a)**

1.9 nm     1.7 nm     1.5 nm

1.3 nm     1.1 nm     1.0 nm

**(b)**

r (nm)

- 1.9 nm
- 1.8 nm
- 1.7 nm
- 1.6 nm
- 1.5 nm
- 1.4 nm
- 1.3 nm
- 1.2 nm
- 1.1 nm
- 1.0 nm

**(c)**

**Large Distances**
**(d > ~ 1.5 nm)**

$r \geq r_0$

Attractive

**Critical Distances**
**(~1.3 nm < d < ~ 1.5 nm)**

$r \leq r_0$

Repulsive

**Small Distances**
**(d < ~ 1.3 nm)**

$r \geq r_0$

Attractive

**Figure 9.**

**TOC graphic:**

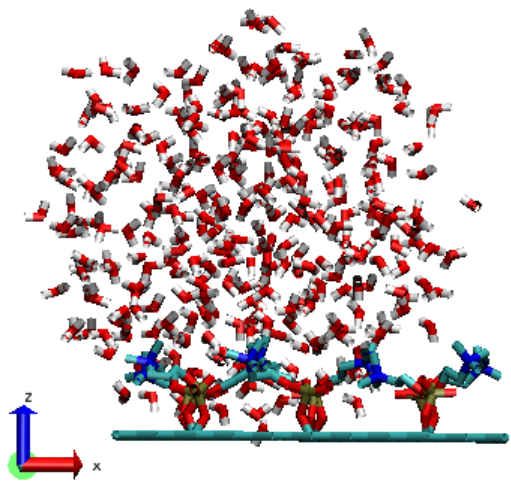